\documentstyle[12pt,moriond,psfig]{article}
\begin{document}
\heading{STAR FORMATION HISTORIES\\ OF LOCAL GROUP DWARF GALAXIES} 

\author{E.K. Grebel $^{1}$} {$^{1}$ UCO/Lick Observatory, University of 
California at Santa Cruz, CA 95064, USA.}

\begin{moriondabstract}
40 galaxies are currently known as, or suspected to be LG members within a
radius of 1.8 Mpc.  37 of these galaxies are dwarf galaxies with $M_B > 
-18.5$ mag.  I present a compilation of the star formation histories of 
dwarf irregulars, dwarf ellipticals, and dwarf spheroidals in the Local Group, 
and visualize their evolutionary histories through Hodge's population boxes.

   The study of these star formation histories is a multi-parameter problem:
Ages, metallicities, population fractions, and spatial variations must be
determined, which depend crucially on the knowledge of reddening and distance.

   Even the lowest-mass dwarf spheroidal galaxies can show significant spatial
variations in star formation, ages, and metallicities.  All Local Group dwarf 
galaxies appear to contain old ($>10$ Gyr) and intermediate-age (1--10 Gyr) 
populations irrespective of morphological type.  No two galaxies are alike.
They vary widely in star formation histories, metallicities, fractions, and 
ages of their subpopulations even within the same morphological type.  Star 
formation has occurred either in distinct episodes or continuously over long 
periods of time.  
\end{moriondabstract}

\section{Importance and uniqueness of the Local Group}

Our knowledge of galaxy formation and evolution starts with the 
study and understanding of our nearest extragalactic neighbours, 
the galaxies in the Local Group (LG).  LG members are close enough
for determinations of ages and metallicities of their resolved 
stellar populations, ISM abundance studies, and detailed 
investigations of their star formation histories.  Galaxies in the
LG comprise a variety of morphological types and cover a wide range of
ages, metallicities, and masses.  The LG contains 
both fairly isolated galaxies as well as galaxies in subgroups,
which enables us to study environmental influences. 
The history of the LG also sheds 
light on the formation and evolution of our own Milky Way.  A thorough
understanding of the processes in the LG, where stellar populations
can be resolved, is the precondition for understanding distant, 
unresolvable galaxies.

\begin{figure}[ht] 
\centerline{\vbox{
\psfig{figure=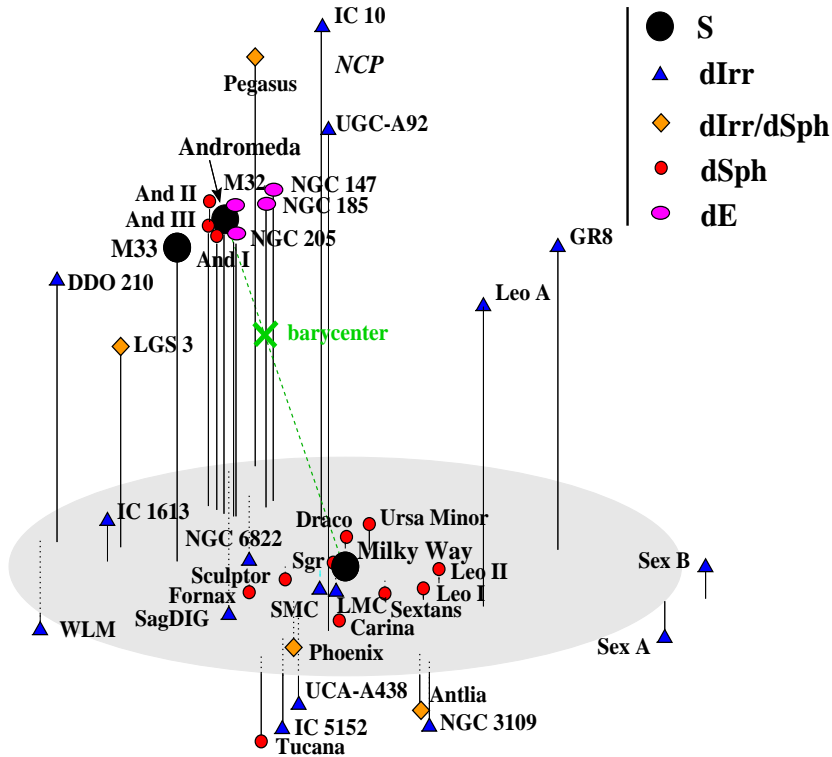,height=10cm,width=11cm,angle=270}
}}
\caption{ \label{fig_LGsketch}
Sketch of the Local Group.  Note the clustered distribution of dEs and
dSphs around the two large spirals, while dIrrs tend to be more distant
and isolated.  The sketch only gives an approximate representation and
is not to scale.}
\end{figure}

\section{The galaxy content of the Local Group}

\subsection{Morphological types}

We distinguish four basic morphological galaxy types in the LG:
spirals (S), dwarf irregulars
(dIrr), dwarf ellipticals (dE), and dwarf spheroidals (dSph).  
The dIrrs are gas-rich, irregularly shaped galaxies with recent or 
ongoing star formation.  The dEs are compact, show
very pronounced, dense, bulge-like cores, and may contain gas. 
They contain mainly old and intermediate-age populations and
show in part recent star formation.  The dSphs are the least luminous, 
least massive galaxies known and, surprisingly, almost devoid of gas. 
They do not have a pronounced 
nucleus and show little central concentration (see also [12]). 
They are dominated by old or intermediate-age populations.

A few galaxies are classified as intermediate between dIrrs and dSphs 
and may evolving from dIrrs to dSphs.  Like dSphs they are dominated
by old populations but contain gas and show some recent star formation.

For the purpose of this review, old refers to ages $>10$ Gyr,
intermediate-age populations range from 1 to 10 Gyr, and young denotes
populations younger than 1 Gyr. 

\begin{figure}[ht] 
\centerline{\vbox{
\psfig{figure=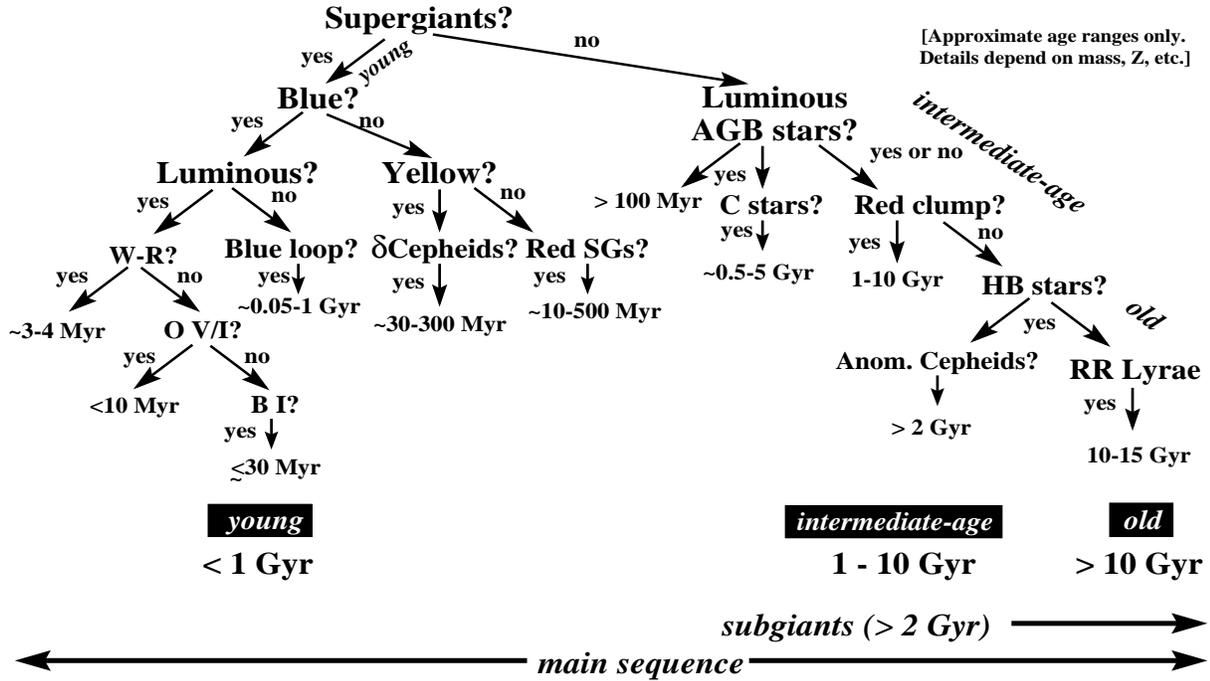,height=9cm,width=16cm,angle=270}
}}
\caption{ \label{fig_starage}
Stars as age indicators for dwarf galaxy studies.  Note how the 
resolution decreases with increasing age.
}
\end{figure}

\subsection{37 dwarfs in the Local Group...}

Grebel [14] derived a zero-velocity surface diameter of $\approx 1.8$ Mpc 
for the LG, which results in 40 galaxies as known or probable members of the 
LG (Fig.\ \ref{fig_LGsketch}).  Only three of these are spirals.  The two 
largest spirals, M\,31 and the Milky Way, contain almost the entire mass in 
the LG.  If we adopt $M_B \ge -18.5$ as cutoff magnitude, all remaining 37 
galaxies qualify as dwarf galaxies.  

The four dEs in the LG are all satellites of 
M\,31.  The 13 dSphs are mostly satellites of the Milky Way or M\,31.  Except 
for the Magellanic Clouds, the 16 dIrrs in the LG tend to be distant and 
isolated galaxies.  Four dwarf galaxies are classified as possible transition
types between dIrrs and dSphs.  None of them is in the immediate vicinity
of a large spiral.  

Recent studies of two dSphs 
uncovered small, fairly young populations (Fornax: [35], 
Sculptor: [9]) indicating a continuous transition
from intermediate types to dSphs.

\subsection{... and counting!}

The full galaxy content and size of the LG are not yet known.  During
the past years, many new faint dwarf members of the LG were discovered.
while other known dwarf galaxies were added to the LG when accurate 
distance measurements became available.  

For instance, the Antlia dwarf
(AM\,1001-270), whose radial velocity $V_{\odot}$ 
[11] and position in the cos\,$\theta$, $V_{\odot}$
diagram [14] suggested possible LG membership was confirmed as 
member through distance determinations by Whiting et al.\ [41]. 
Similarly, the redetermination of the distance
to the LG suspect Leo\,A makes it now a certain member [38].

More sensitive surveys for low-surface-brightness galaxies as well as 
surveys of highly extincted areas of the sky may to uncover additional
LG members (e.g., [23]).  These surveys will
most likely contribute galaxies at the faint end of the luminosity function.
For recent compilations and membership discussions see [39] 
and [14].

\section{Deriving star formation histories}

\subsection{Stellar population studies}

Studies of the evolutionary history of a galaxy usually combine the use of 
deep CCD photometry to derive color-magnitude diagrams (CMDs), and the
use of stars as age tracers (Fig.\ \ref{fig_starage}).  

Special types of stars can be crucial for uncovering subpopulations
even when high-quality CMDs are available.  A few examples:
The presence of a small intermediate-age population may not be obvious from 
the CMD of an old dSph, but the detection of carbon stars
traces an intermediate-age population unambiguously.  Anomalous Cepheids
may either be single intermediate-age stars or old binary systems [4], 
thus do not trace intermediate-age populations unambiguously.
The presence of RR Lyrae
stars or blue horizontal branch stars is a certain sign of an old population, 
but the lack thereof does not necessarily imply the absence of an old 
population since second-parameter effects can also play a role.   

The determination of the star formation history of a galaxy is a 
multi-parameter problem.  The interpretation depends crucially on the
knowledge of reddening (both foreground and internal extinction) and distance.  
In addition, data may be compromised by crowding and high or variable
internal extinction in gas-rich galaxies.  With increasing distance it 
becomes more and more difficult to study the properties of the older 
populations.  Mixed populations and the age-metallicity-reddening-distance
degeneracy present major challenges and may lead
to ambiguous interpretations.  The ambiguity can be reduced by making use
of all available information, e.g., by deriving the foreground extinction
from reddening maps (e.g., [5]), or by using spectroscopic
abundances if available.  

Synthetic CMDs, sophisticated modelling techniques and statistical
evaluations allow it to decrease the ambiguities to some extent and to extract 
detailed star formation histories (e.g., [36], [16], 
[13], [37], [10], and [21]). 

\begin{figure}[bt] 
\centerline{\vbox{
\psfig{figure=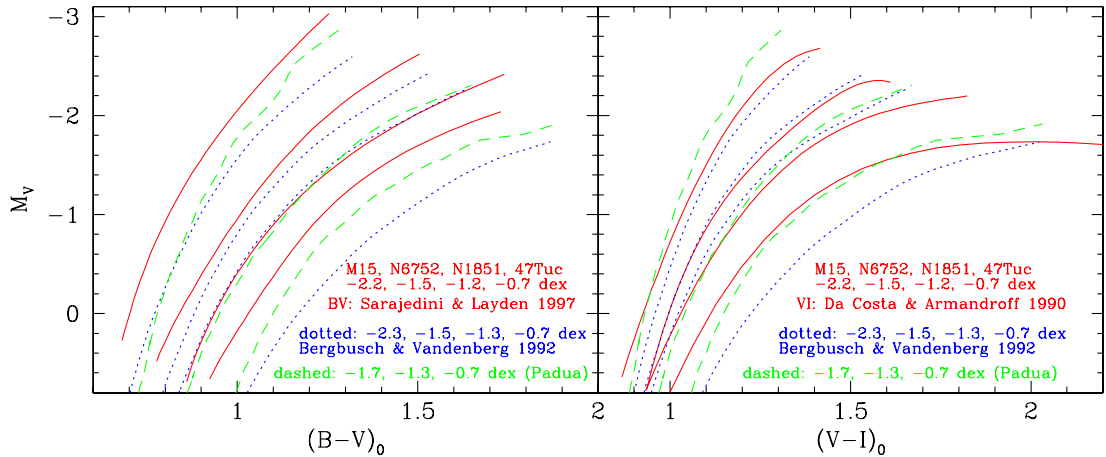,height=7cm,width=16cm,angle=270}
}}
\vspace{-0.5cm}
\caption{ \label{fig_fiduc}
Comparison of fiducial lines of the observed red giant branches of Galactic 
globular clusters from [30] and [7] with theoretical isochrones for similar 
abundances from [2] and [3].
}
\end{figure}

These analyses depend on how well evolutionary models reproduce observational
parameters (Fig.\ \ref{fig_fiduc}) as well as on assumptions about 
IMFs and metallicity evolution.  Results obtained with different evolutionary
models may not be directly comparable and tend to differ in ages, age spreads,
and enrichment.  

\subsection{Other ingredients}

While photometry is the primary method for stellar population studies, 
spectroscopy provides accurate abundances, wind properties, spectral types, 
radial velocities, or stellar velocity dispersions.
The new 10-m-class telescopes will make many more 
LG galaxies accessible for spectroscopic studies and help to uncover 
enrichment histories and chemical evolution (as well as internal kinematics
and dark matter content).

Proper motions are currently available only for a few nearby dwarfs --
Sagittarius [22], LMC \& SMC [24], Draco \& Ursa Minor [32], and Sculptor [33].
The planned satellite missions will increase and improve astrometric 
information for LG galaxies and help to derive their kinematics and 
orbits in the LG.  Ultimately this may help to constrain past interaction
events.

Star formation histories must be supplemented by studies of the ISM in and
around galaxies -- gas content and distribution, current star formation
rates, kinematics, abundances (e.g., [42]).
These studies help to understand the impact of 
accretion of gas clouds on star formation
(e.g., [20]) and the mystery of the
missing gas in dSphs [6].   

\section{Spatial variations in the star formation history}

Not only do most well-studied dwarf galaxies show evidence
for several episodes of star formation, but
star formation and enrichment history can vary with position in a galaxy 
even if this galaxy is a low-mass dwarf galaxy.  
While HST imaging results in excellent resolution and depth, large-area
coverage is needed to map out spatial variations in the evolutionary history.
Many dwarf galaxies 
have extended old stellar ``halos'' (e.g., the dIrrs WLM [29] and Sextans
A [19]; the dE NGC\,147 [17], the dSph/dIrr Antlia ([1], [31]);
the dSphs And\,I [8], Carina [28], and Fornax ([14], [35]).  

\subsection{Example Fornax}

The predominantly intermediate-age dSph Fornax is a good example for 
pronounced spatial variations
in age and metallicity.  Star formation lasted up to 6 Gyr longer in the
central regions than in the outer areas, and the slopes of the
red giant branches indicate a pronounced metallicity gradient [14].  
In addition the central regions show a large metallicity spread of more
than 1 dex, while the outer areas have a mean metallicity of $-1.3$ dex
with a spread of $\approx 0.3$ dex.  Subpopulations distinct in metallicity
are visible in the metallicity distribution functions derived from red
giants [15].   
Fornax's populations have different centroids [35].    
The most metal-poor globular clusters in Fornax have the largest projected 
galactocentric distances.  The two metal-poor ($\approx -2$ dex)
clusters GC\,\#5 and GC\,\#1 exhibit the second-parameter effect [34].

\section{Visualizing evolutionary histories through population boxes}

\begin{figure}[ht] 
\centerline{\vbox{
\psfig{figure=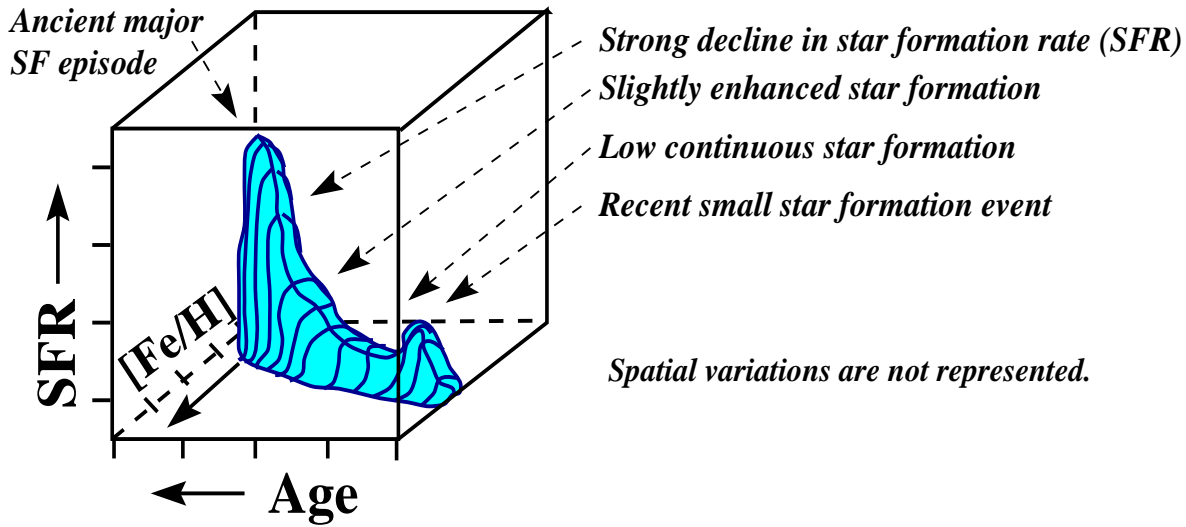,height=7cm,width=16cm,angle=270}
}}
\caption{ \label{fig_popbox}
A sample population box for a fictitious galaxy. }
\end{figure}

\begin{figure}[ht] 
\centerline{\vbox{
\psfig{figure=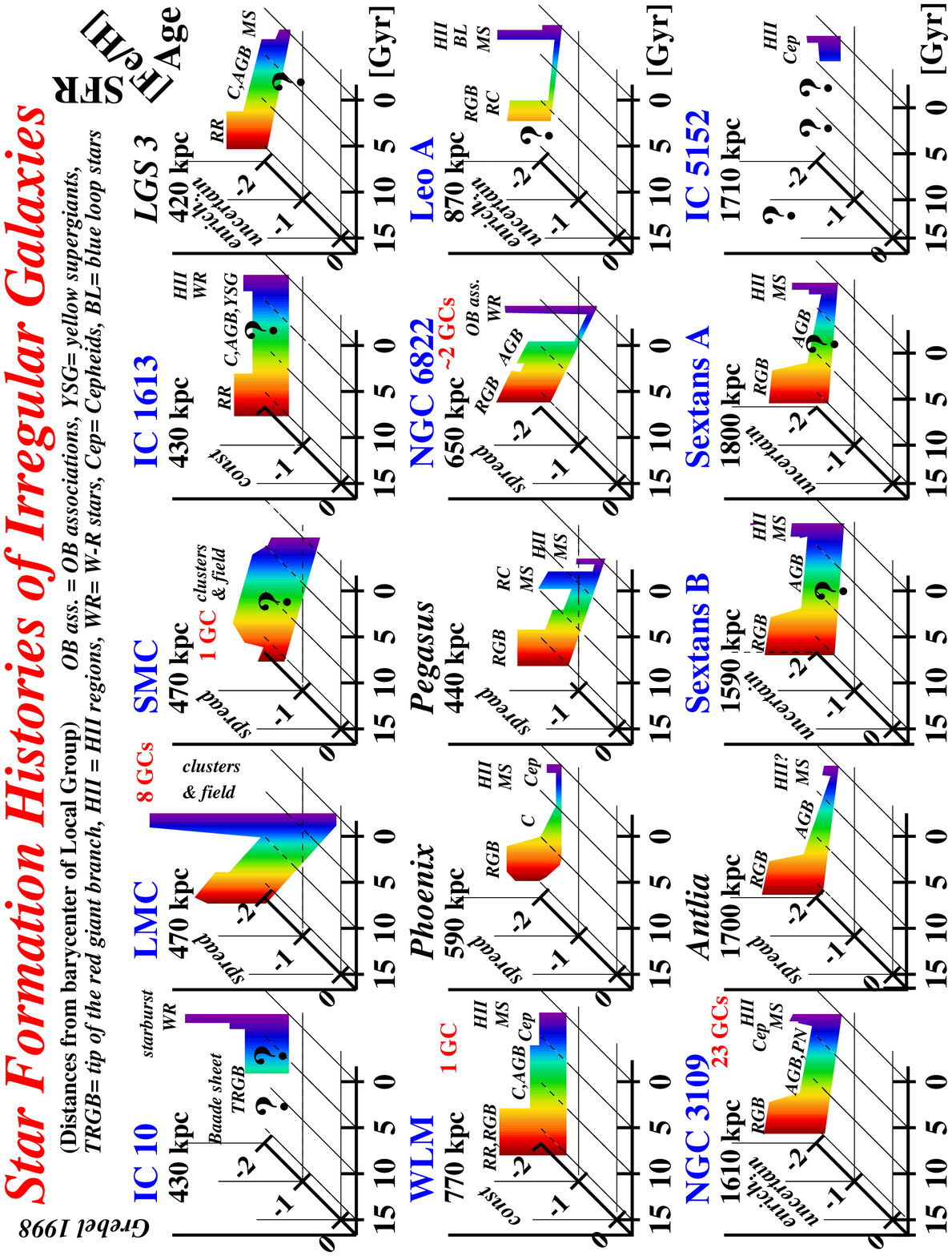,height=24cm,width=16cm,angle=0}
}}
\caption{ \label{fig_SFR_dI}
Population boxes for LG dIrr (Roman fonts) and and dIrr/dSph
galaxies (italics).}
\end{figure}

\begin{figure}[ht] 
\centerline{\vbox{
\psfig{figure=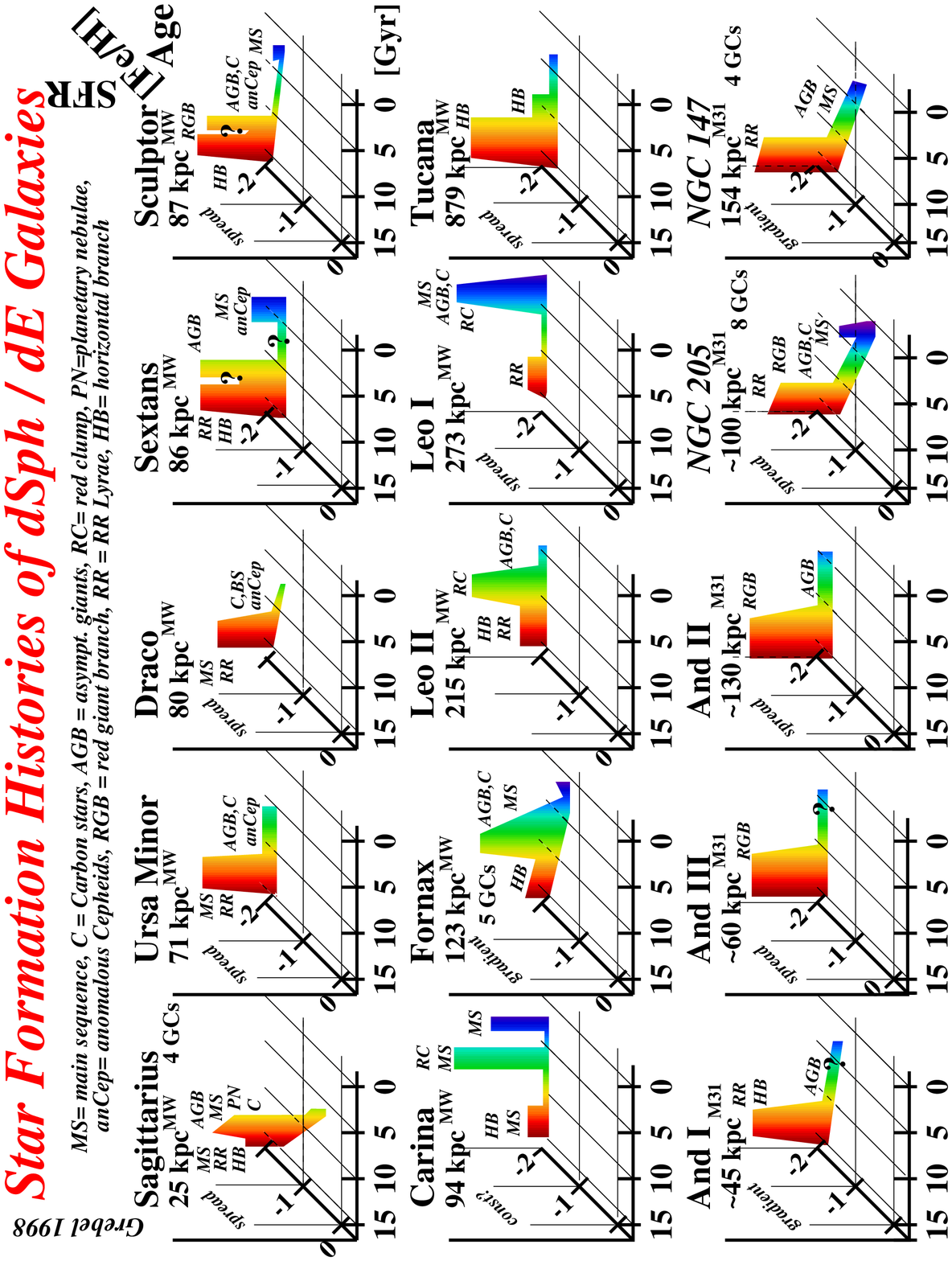,height=24cm,width=16cm,angle=0}
}}
\caption{ \label{fig_SFR_dS}
Population boxes for LG dSph (Roman fonts) and dE galaxies 
(italics). }
\end{figure}

Hodge [18] introduced population boxes as an instructive 
3-D visualization of the
evolutionary history of a galaxy (Fig.\ \ref{fig_popbox}). 
In Figs.\ \ref{fig_SFR_dI} and \ref{fig_SFR_dS} I am presenting an updated
version of the compilation of star formation histories of LG galaxies by
Grebel ([14]; see Tab.\ 4 \& 5 therein for a compilation of ages,
populations, metallicities, and other properties).  The current 
update is based on all available information on the star formation histories
and metallicities in LG galaxies as of May 1998.  As before it combines data
from heterogeneous sources obtained using different observational techniques
and different theoretical models.  

Data on metallicities and enrichment
are often uncertain.  In many cases only two data points were available from
the literature, which I assigned to their respective population and 
connected by a straight line.  There is of course no evidence that the 
enrichment did actually proceed smoothly and linearly.  In many cases
substantial metallicity spreads are found already among the old populations, 
while other galaxies show very little evidence of metallicity spreads or
enrichment.  A galaxy that experienced hardly any enrichment
despite repeated star formation episodes appears to be Carina.  

Time-dependent star formation rates (SFR) are not yet known
for most galaxies.  Therefore I use qualitative estimates of subpopulation
ratios and do not label the SFR axis.  Since the current diagrams use a
linear time scale as opposed to the logarithmic scale used previously
I would like to emphasize that the time resolution decreases with age, and that
we have very little information about details of star formation episodes
longer ago than 10 Gyr.  Many galaxies seem to be more active at the present
time than within the past few Gyr, which may be in part a selection effect
since present-day activity can be easily 
traced by H{\sc ii} regions.

Both internal and external processes determine a galaxy's evolution and gas
loss.  For the 9 dSphs associated with the Milky Way (panels 1 and 
2 in Fig.\ \ref{fig_SFR_dS} the dominant populations tend to become younger
with increasing distance from the Milky Way as 
suggested by van den Bergh [40] in his tidal stripping/ram pressure
scenario.  

The initial star formation episode 
may cause much of the star-forming material to be expelled if the galaxy
has a low mass, while more massive galaxies are able to retain their
gas much longer or even continue to form stars over a Hubble time and
more.  
Detailed modeling of observed color-magnitude diagrams shows that star
formation has often been continuous over long periods of time in the more
massive dIrrs (e.g., [13]), interrupted by short periods of inactivity. 
The observed star formation histories and gas content
of low-mass galaxies appear to support
that dIrrs may eventually evolve into dSphs.

The detection of small, relatively young populations in dSphs
with extremely low H{\sc i} content (e.g., Fornax [35])
indicates that dSphs may be able 
to retain or regain part of the gas that may have been expelled
earlier during star formation episodes.  Carignan's (this conference)
exciting discovery of an H{\sc i} ring around the 
Sculptor dSph indicates that expelled gas may remain with its
parent galaxy and that ram pressure or tidal stripping did not suffice
to strip the galaxy entirely of its gas.  Are more dSphs embedded in 
rings or spheres of expelled gas and may have future minor star formation
events?

\section{Summary and Outlook}

All LG galaxies vary widely in\vspace{-0.3cm} 
\begin{itemize}{}\parsep=0pt\itemsep=-0.5pt\parskip=0pt
\item their star formation histories, 
\item their metallicities and enrichment, and 
\item fractions and ages of their subpopulations,  
\end{itemize}
\vspace{-0.3cm}
even within the same morphological 
type.  No two dwarf galaxies are alike.

Many LG galaxies show complex star formation histories even within their
older populations and may exhibit preserved spatial variations in age
and metallicity.  Star formation tends to continue longer in the central
regions.   

All galaxies studied in sufficient detail were found to contain old
($> 10$ Gyr) and intermediate-age ($1-10$ Gyr) populations irrespective
of morphological type.
Even many dIrr galaxies show sizeable fractions of old populations.

Main areas for future study include the determination of accurate distances
to LG members, accurate age determinations for subpopulations, spectroscopic
abundance determinations and studies of the chemical enrichment history,
and studies of spatial variations.  With accurate astrometric data we will
be able to place the individual galaxies in the context of the
evolution of the LG as a whole.  An intriguing question is whether 
more accurate proper motions and space motions will verify that the
Milky Way satellites are moving along two polar great circles or orbital
planes (see [25], [26], [27]) 
around the Milky Way.  Could the dwarfs we observe today be the few survivors 
on polar orbits, while the early galaxy population of the LG would have been
substantially more numerous and since have been accreted?
  
\subsection*{Acknowledgments}

I thank Dennis Zaritsky for support through NASA LTSA grant NAG-5-3501.


%
\vfill
\end{document}